\documentclass[preprint]{aastex}

\begin{document}

\shortauthors{Gordon et al.\ }
\shorttitle{The Flux Ratio Method}

\title{The Flux Ratio Method for Determining the Dust Attenuation of
Starburst Galaxies}

\author{Karl D.\ Gordon\altaffilmark{1}, 
Geoffrey C.\ Clayton\altaffilmark{2},
Adolf N.\ Witt\altaffilmark{3}, and
K.\ A.\ Misselt\altaffilmark{2}}
\altaffiltext{1}{Steward Observatory, University of Arizona,
   Tucson, AZ 85721; kgordon@as.arizona.edu}
\altaffiltext{2}{Department of Physics \& Astronomy, Louisiana State
   University, Baton Rouge, LA 70803; (gclayton,misselt)@fenway.phys.lsu.edu}
\altaffiltext{3}{Ritter Astrophysical Research Center, The University
   of Toledo, Toledo, OH 43606; awitt@dusty.astro.utoledo.edu}

\begin{abstract}
The presence of dust in starburst galaxies complicates the study of
their stellar populations as the dust's effects are similar to those
associated with changes in the galaxies' stellar age and metallicity.
This degeneracy can be overcome for starburst galaxies if
UV/optical/near-infrared observations are combined with far-infrared
observations.  We present the calibration of the flux ratio method for
calculating the dust attenuation at a particular wavelength,
$Att(\lambda)$, based on the measurement of $F(IR)/F(\lambda)$ flux
ratio.  Our calibration is based on spectral energy distributions
(SEDs) from the PEGASE stellar evolutionary synthesis model and the
effects of dust (absorption and scattering) as calculated from our
Monte Carlo radiative transfer model.  We tested the attenuations
predicted from this method for the Balmer emission lines of a sample
starburst galaxies against those calculated using radio observations
and found good agreement.  The UV attenuation curves for a handful of
starburst galaxies were calculated using the flux ratio method, and they 
compare favorably with past work.  The relationship between
$Att(\lambda)$ and $F(IR)/F(\lambda)$ is almost completely independent
of the assumed dust properties (grain type, distribution, and
clumpiness).  For the UV, the relationship is also independent of the
assumed stellar properties (age, metallicity, etc) accept for the case
of very old burst populations.  However at longer wavelengths, the
relationship is dependent on the assumed stellar properties.
\end{abstract}

\keywords{galaxies: ISM -- galaxies: starburst}

\section{Introduction}

To study galaxies, it is crucial to be able to separate the effects of
the dust intrinsic to the galaxy from those associated with the
galaxy's stellar age and metallicity.  Currently, the accuracy of
separating the stars and dust in galaxies is fairly poor and the study
of galaxies has suffered.  This is in contrast with studies of individual
stars and their associated sightlines in the Milky Way and nearby
galaxies for which the standard pair method \citep{mas83} works quite
well at determining the effects of dust on the star's spectral energy
distribution (SED).  The standard pair method is based on comparing a
reddened star's SED with the SED of an unreddened star with the same
spectral type.  Application of the standard pair method to galaxies is
not possible as each galaxy is the result of a unique evolutionary
history and, thus, each has a unique mix of stellar populations and
star/gas/dust geometry.

Nevertheless, it would be very advantageous to find a method which
would allow one to determine the dust attenuation of an individual
galaxy.  Such a method would greatly improve the accuracy of different
star formation rate measurements.  For example, two widely used star
formation rate measurements are based on UV and H$\alpha$
luminosities.  Both are affected by dust and this limits their
accuracy \citep{ken98,sch99}.  The importance of correcting for the
effects of dust in galaxies has gained attention through recent
investigations into the redshift dependence of the global star
formation rate \citep{mad98, ste99}.  The uncertainty in the
correction for dust currently dominates the uncertainty in the
inferred star formation rate in galaxies
\citep{pet98, meu99} and conclusions about the evolution of galaxies
\citep{cal99}.

Initially, the effects of dust in galaxies were removed using a screen
geometry.  This assumption has been shown to be a dangerous
oversimplification as the dust in galaxies is mixed with the stars.
Radiative transfer studies have shown that mixing the emitting sources
and dust and having a clumpy dust distribution produces highly
unscreen-like effects \citep{wit92, wit96, gor97, fer99, tak99,
wit99}.  For example, the traditional reddening arrows in color-color
plots turn into complex, non-linear reddening trajectories.  In
general, the attenuation curve of a galaxy is not directly
proportional to the dust extinction curve and its shape changes as a
function of dust column (e.g., Figs.~6 \& 7 of \citet{wit99}).

While the various radiative transfer studies have made it abundantly
clear that correcting for the effects of dust in galaxies is hard,
none have come up with a method that is not highly dependent on the
assumed dust grain characteristics, star/gas/dust geometry, and
clumpiness of the dust distribution.  This has led to a search for
empirical methods.  For galaxies with hydrogen emission lines, it is
possible to determine the slope and, with radio observations, the
strength of the galaxies' attenuation curves at the emission line
wavelengths \citep{cal94, smi95}.  Unfortunately, this method is
limited to the select few wavelengths associated with hydrogen
emission lines.  In the pioneering study of the IUE sample of
starburst galaxies \citep{kin93}, \citet{cal94} used a variant of the
standard reddened star/unreddened star method to compute the average
attenuation curve for these galaxies.  This work binned the sample
using $E(B-V)$ values derived from the H$\alpha$ and H$\beta$ emission
lines and assigned the lowest $E(B-V)$ bin the status of unreddened.
While this work was a significant advance in the study of dust in
galaxies, it is only applicable to statistical studies of similar
samples of starburst galaxies, not individual galaxies \citep{saw98}.

More recently, \citet{meu99} derived a relationship between the slope
of the UV spectrum of a starburst galaxy and the attenuation suffered
at 1600~\AA, $Att(1600)$, using the properties of the IUE sample.
This slope is parameterized by $\beta$ where the UV spectrum is fit to
a power law ($F(\lambda) \propto \lambda^{-\beta}$) in the wavelength
range between 1200 and 2600~\AA\ \citep{cal94}.  The purpose of
\citet{meu99} was to calculate the attenuation suffered by high
redshift starburst galaxies using only their UV observations.  From
our radiative transfer work, we have found that this relationship is
strongly dependent on the star/gas/dust geometry, dust grain
properties, and dust clumpiness \citep{wit96, wit99} as suspected by
\citet{meu99}.  Fig.~11 of \cite{wit99} shows the dependence of
$Att(1600)$ on $\Delta\beta$ ($= \beta - 2.5$) for various geometries,
dust clumpinesses, and dust types.  \citet{meu99} used the observed
relationship between $F(IR)/F(1600)$ and $\beta$ for starburst
galaxies, combined with a semi-empirical calibration between
$F(IR)/F(1600)$ and $Att(1600)$ to determine the relationship between
$Att(1600)$ and $\beta$.  The correlation between $F(IR)/F(UV)$ and
$\beta$ was first introduced by \citet{meu97} where $F(UV) = F(2200)$.
\citet{wit99} discovered that the relationship between $F(IR)/F(1600)$
and $Att(1600)$ was almost completely independent of the star/gas/dust
geometry, dust grain properties, and dust clumpiness (see Fig.~12b of
\citet{wit99}).  This implies that $F(IR)/F(1600)$ is a much better
indicator of $Att(1600)$ than $\beta$.

This opened the possibility that the $F(IR)/F(\lambda)$ might be a
good measure of $Att(\lambda)$ and was the motivation for this paper.
Qualitatively, there is good reason to think that a measure based on
the flux at a wavelength $\lambda$ and the total flux absorbed and
re-emitted by
dust, $F(IR)$, should be a measure of $Att(\lambda)$.  This is
basically a statement of conservation of energy.  Evidence that
$F(IR)/F(UV)$ is a rough indicator of $Att(UV)$ in disk galaxies is
given by \citet{wan96}.  The details of the relationship between
$F(IR)/F(\lambda)$ and $Att(\lambda)$ will be dependent on the
stellar, gas, and dust properties of a galaxy.  Thus, a calibration of
the relationship is necessary.

In \S\ref{sec_method}, we calibrate the relationship between
$F(IR)/F(\lambda)$ and $Att(\lambda)$ for UV, optical, and near-IR
wavelengths using a stellar evolutionary synthesis model combined with
our dust radiative transfer model.  This allowed us to investigate the
dependence of the relationship on stellar parameters (age, star
formation type, and metallicity) and dust parameters (geometry, local
dust distribution, dust type, and the fraction of Lyman continuum
photons absorbed by dust).  We show a comparison of $Att(H\alpha)$,
$Att(H\beta)$, and $Att(H\gamma)$ values determined with this flux
ratio method and the radio method \citep{con92} for 10 starburst
galaxies in \S\ref{sec_compare}.  In \S\ref{sec_app}, we apply
the flux ratio method to construct the UV attenuation curves for 8
starburst galaxies.  The implications this work are discussed in
\S\ref{sec_discuss}.

\section{The Flux Ratio Method \label{sec_method}}

\subsection{$F(IR)/F(\lambda)$ Flux Ratio}

In a galaxy, almost all of the photons absorbed by dust are emitted by
stars and gas in the UV, optical, and near-IR.  This energy heats the
dust which then re-emits in the mid- and far-infrared (small and large 
dust grains).  Thus, the
ratio of the total infrared flux to the flux at a particular
wavelength is
\begin{equation}
\frac{F(IR)}{F(\lambda)} = \frac{a_d F(LyC) + (1 - a_d)F(Ly\alpha) +
   \int_{912~\AA}^\infty f(\lambda',0) \left(1 - C(\lambda') \right)
   d\lambda'}{\lambda f(\lambda,0) C(\lambda)}
\label{eq_ratio}
\end{equation}
where $F(IR)$ is the total IR flux in ergs cm$^{-2}$ s$^{-1}$,
$F(LyC)$ is the total unattenuated stellar flux below 912~\AA\ in ergs
cm$^{-2}$ s$^{-1}$, $a_{d}$ is the fraction of $F(LyC)$ absorbed by
dust internal to the \ion{H}{2} regions \citep{pet72, mat86},
$F(Ly\alpha)$ is the $Ly\alpha$ emission line flux, $f(\lambda,0)$ is
the unattenuated stellar/nebular flux in ergs cm$^{-2}$ s$^{-1}$
\AA$^{-1}$, $C(\lambda) = 10^{-0.4Att(\lambda)}$, and $Att(\lambda)$ is
the attenuation at $\lambda$ in magnitudes.  For emission lines, the
denominator of eq.~\ref{eq_ratio} becomes
$(1-a_d)F(\lambda,0)C(\lambda)$ where $F(\lambda,0)$ is the intrinsic
integrated flux of the emission line.  The $Ly\alpha$ line is
resonantly scattered and, thus, is completely absorbed by the dust
internal to the \ion{H}{2} regions.  Eq.~\ref{eq_ratio} is similar to
eq.~3 of \citet{meu99}, but includes an additional term to account for
the Lyman continuum photons absorbed by dust.

\subsection{Relationship between $F(IR)/F(\lambda)$ and $Att(\lambda)$}

We can calculate the relationship between $F(IR)/F(\lambda)$ and
$Att(\lambda)$ by using a stellar evolutionary synthesis (SES) model
and a dust radiative transfer model.  We use the PEGASE SES model
\citep{fio97, fio99} which gives the SEDs of stellar populations with
a range of ages, type of star formation (burst/constant), and
metallicity.  One strength of the PEGASE model is that it computes the
continuum and emission lines expected from gas emission as well as the
stellar emission.  We used a Salpeter IMF for the PEGASE calculations.
The SES model SEDs give $F(LyC)$, $f(\lambda,0)$, and emission line
$F(\lambda,0)$ values.  The effects of dust were calculated using the
DIRTY radiative transfer model \citep{wit99}.  The DIRTY model gives
the attenuation curves, $Att(\lambda)$, for a range of spherical
star/gas/dust global geometries (shell, dusty, or cloudy), local dust
distribution (homogeneous or clumpy), Milky Way \citep{car89} or Small
Magellanic Cloud \citep{gor98} dust grain characteristics, and dust
columns ($\tau_V = 0.25 - 50$).  The cloudy geometry has dust
extending to 0.69 of the system radius and stars extending to the
model radius.  The dusty geometry has both dust and stars extending
to the model radius.  This geometry represents a uniform mixture of
stars and dust.  The shell geometry has stars extending to 0.3 of the
model radius and dust extending from 0.3 to 1 of the model radius.
These three star/gas/dust geometries are shown pictorially in Figure~1
of \cite{wit99}.  Additional details of the the DIRTY model
calculations can be found in \citet{wit99}.

\begin{figure}[tbp]
\plottwo{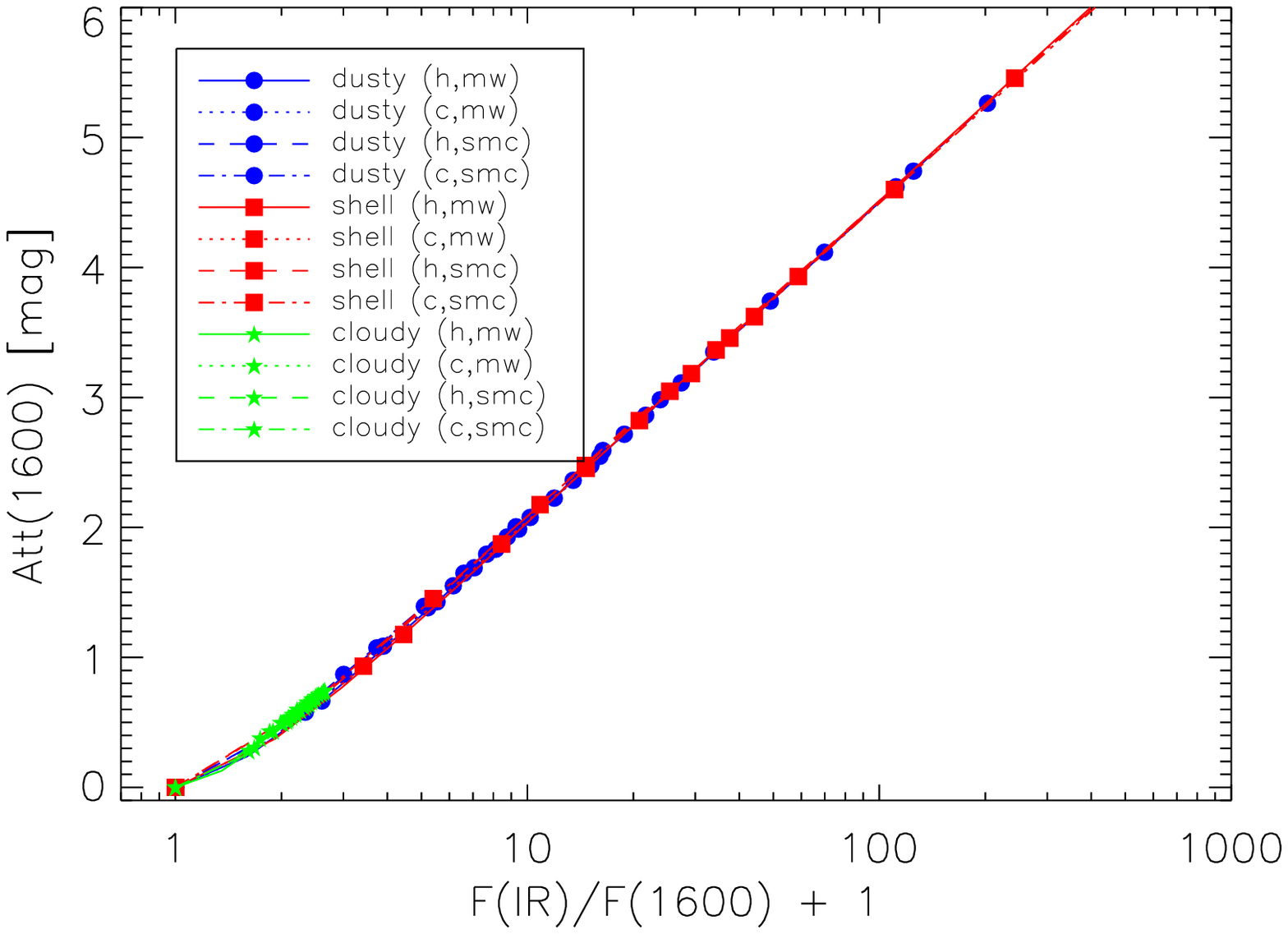}{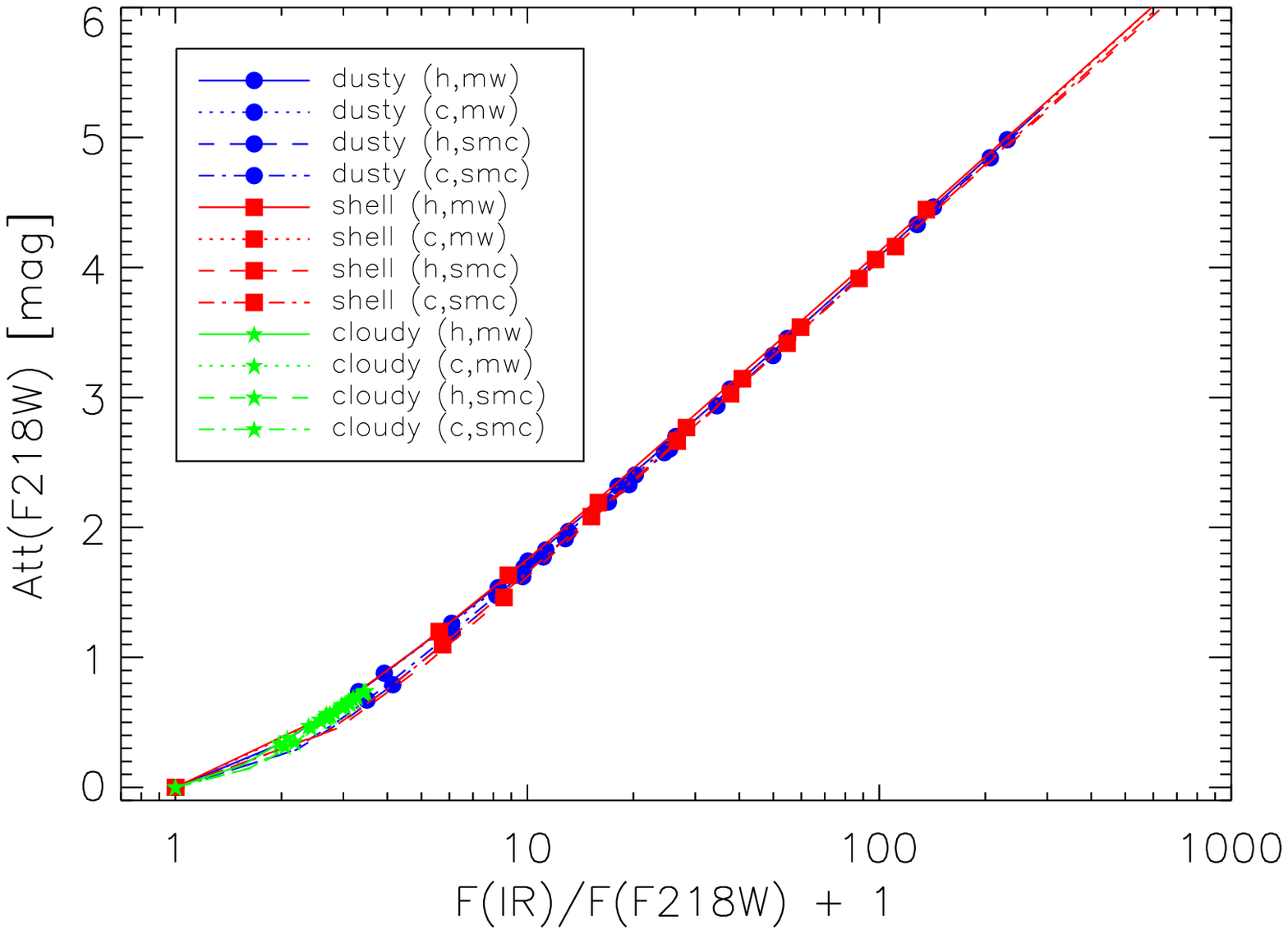} \\
\plottwo{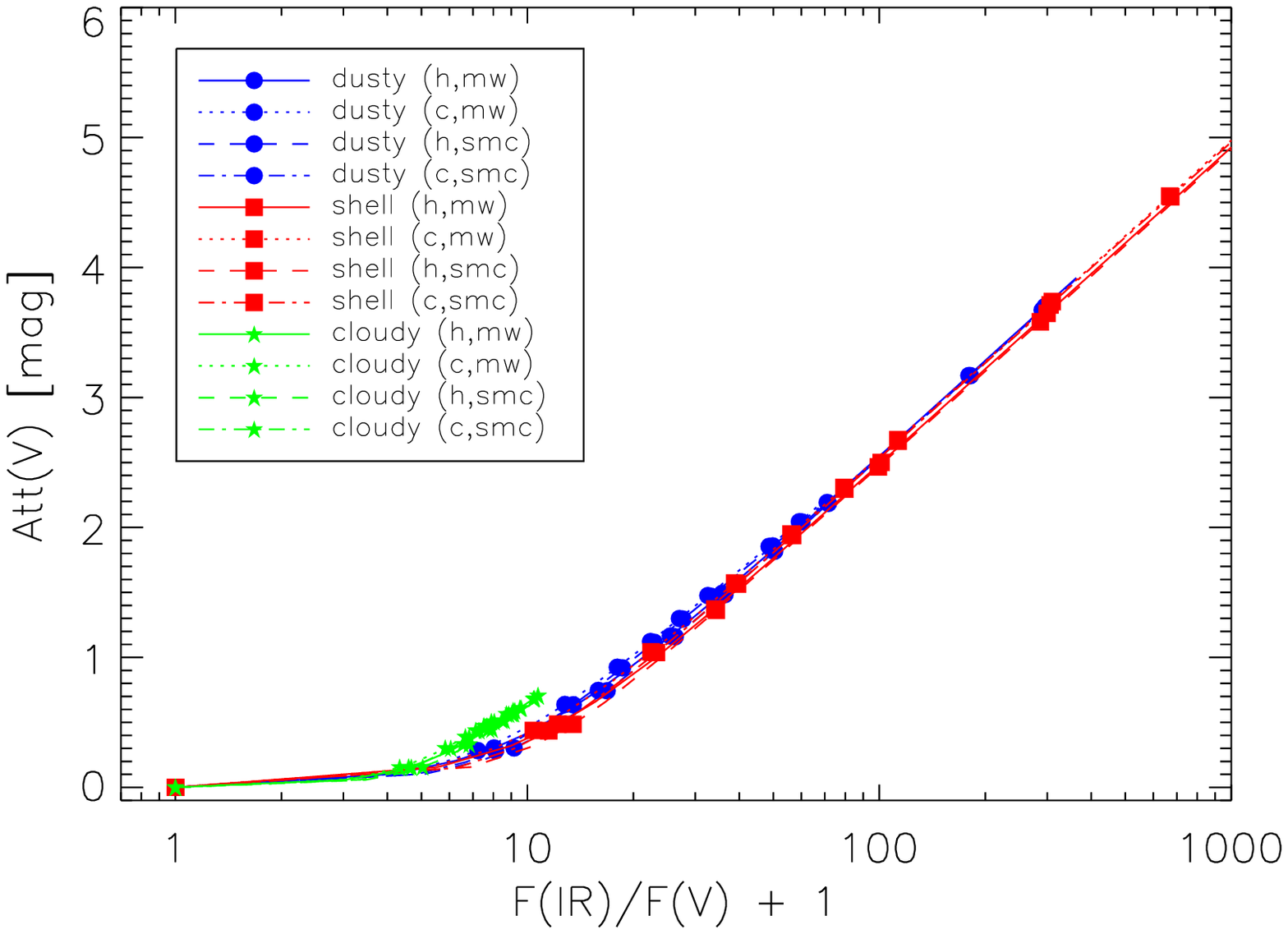}{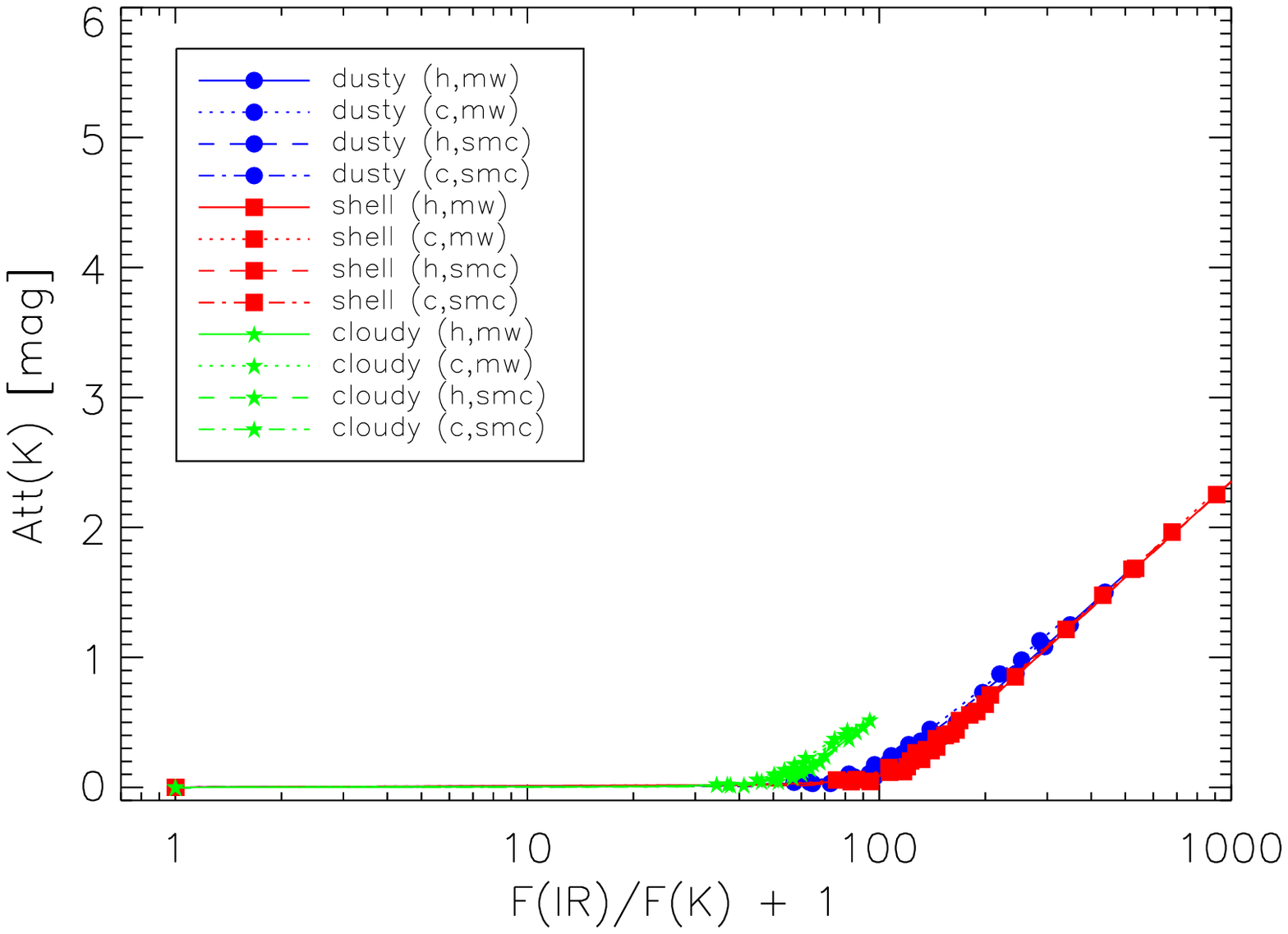}
\caption{The relationship between $F(IR)/F(\lambda)$ and
$Att(\lambda)$ is plotted for the (a) 1600~\AA, (b) HST/WFPC2 F218W,
(c) V, and (d) K bands for all the DIRTY model parameters.  The
parameters of the DIRTY radiative transfer model curves include three
star/gas/dust geometries (cloudy, dusty, shell), two dust grain
properties (MW, SMC), and two dust distributions (h = homogeneous, c =
clumpy).  The details of these parameters can be found in
\citet{wit99}. \label{fig_fr_one}}
\end{figure}

In Figure~\ref{fig_fr_one}, we plot the the relationship between
$F(IR)/F(\lambda)$ and $Att(\lambda)$ for the \citet{meu99} 1600~\AA,
HST/WFPC2 F218W, V, and K bands assuming a constant star formation, 10
Myr old, solar metallicity SED, $a_d = 0.25$, and the full range of
dust parameters (see above).  The most surprising result is that this
relationship is not sensitive to the type of dust (MW/SMC) or the
local dust distribution (homogeneous/clumpy).  This is true not just
for the four bands plotted in Fig.~\ref{fig_fr_one}, but for all the
ultraviolet, optical, and near-infrared.  Less surprising is that this
relationship in the V and K bands is sensitive to the presence of
stars outside the dust.  The dusty and shell geometries follow similar
curves while the cloudy geometry follows a different curve.  For the
cloudy geometry, as the attenuation is increased the dominance of the
band flux from the stars outside the dust increases to the point where
the band flux no longer depends on the attenuation (i.e.\ the flux
from the stars attenuated by dust is much less than the flux from the
unattenuated stars).  This is not the case for the dusty and shell
geometries where band flux continues to decrease with increasing
attenuation since all the stars are inside the dust and attenuated to
some degree.

\begin{figure}[tbp]
\plottwo{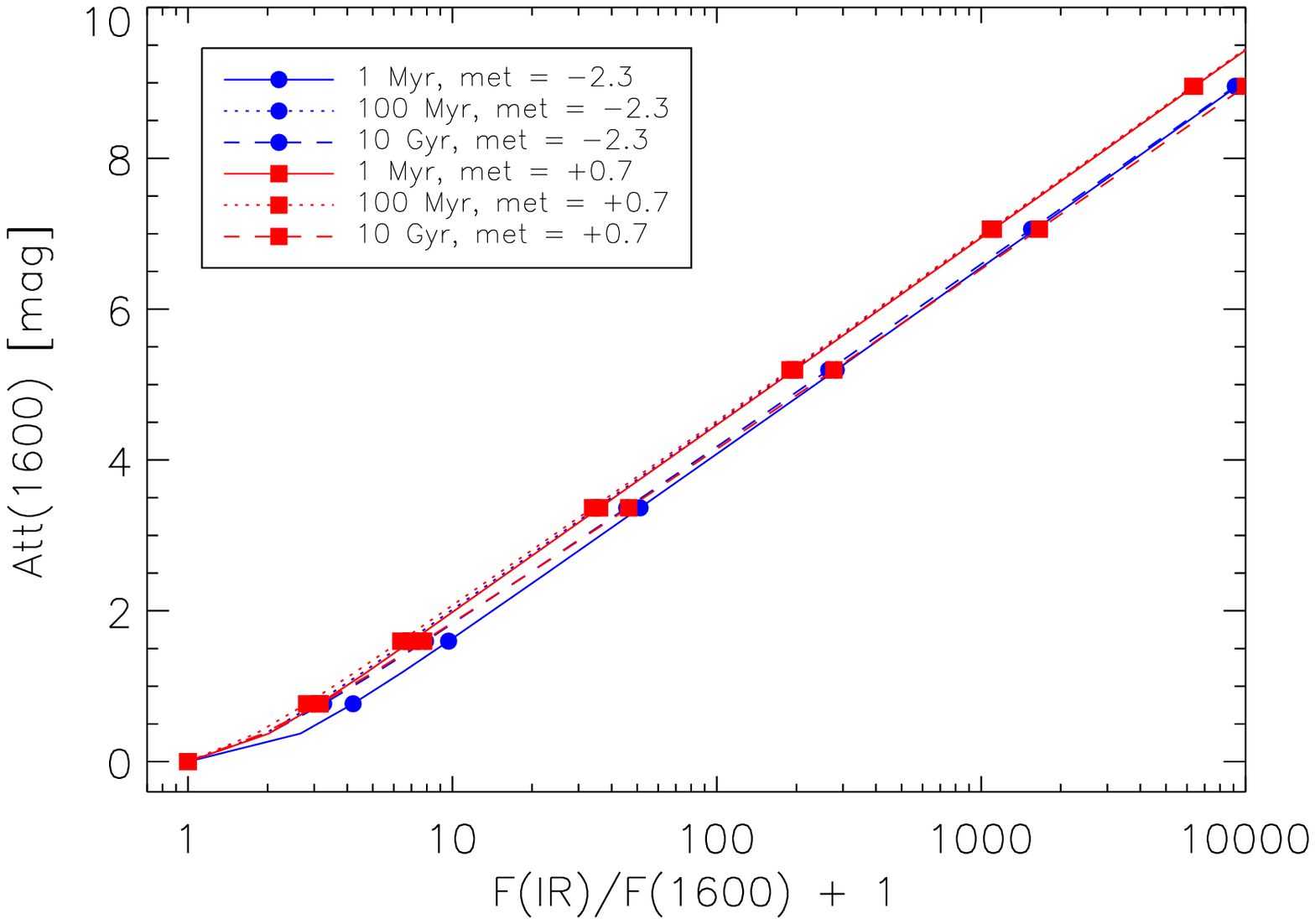}{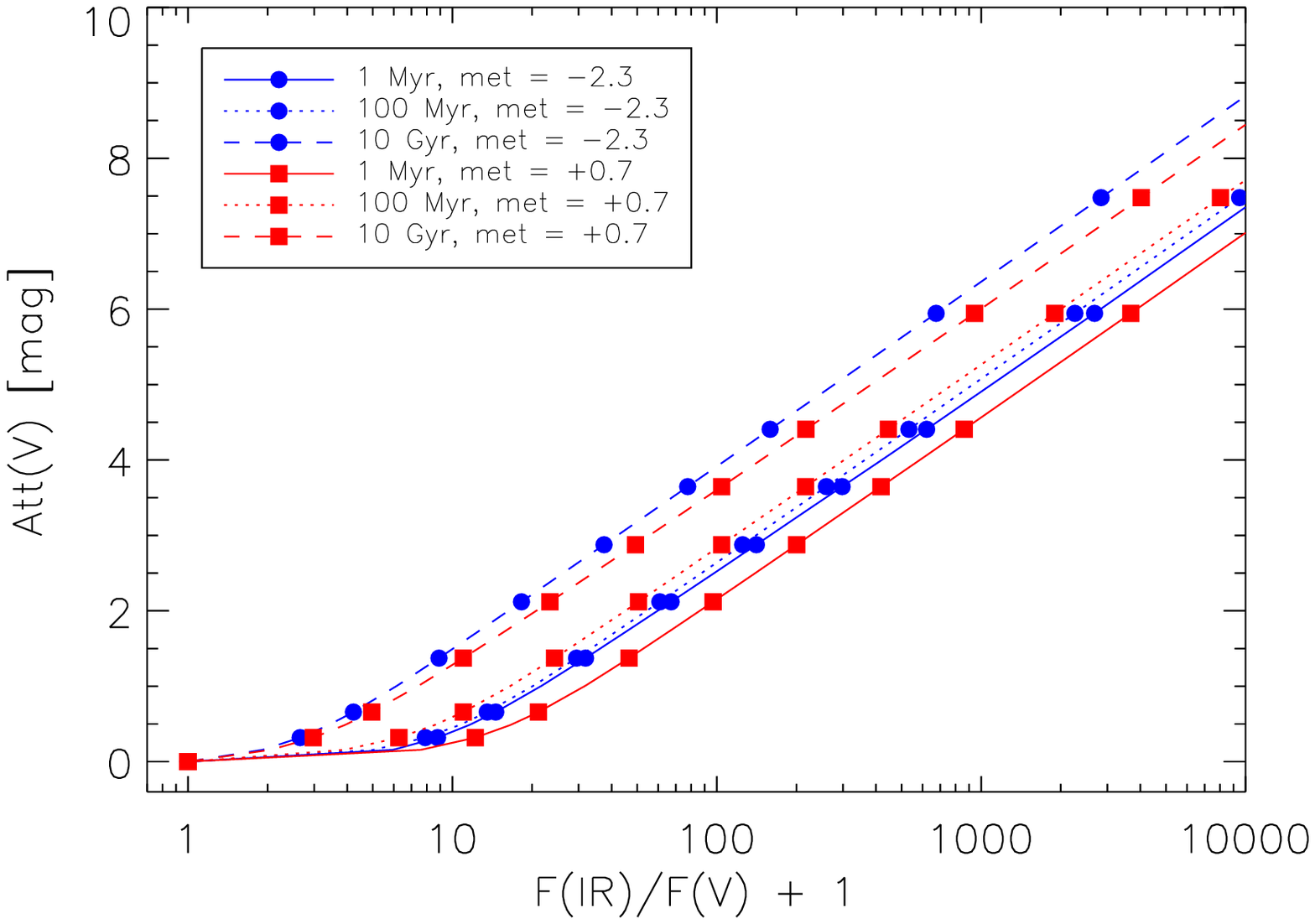} \\
\plottwo{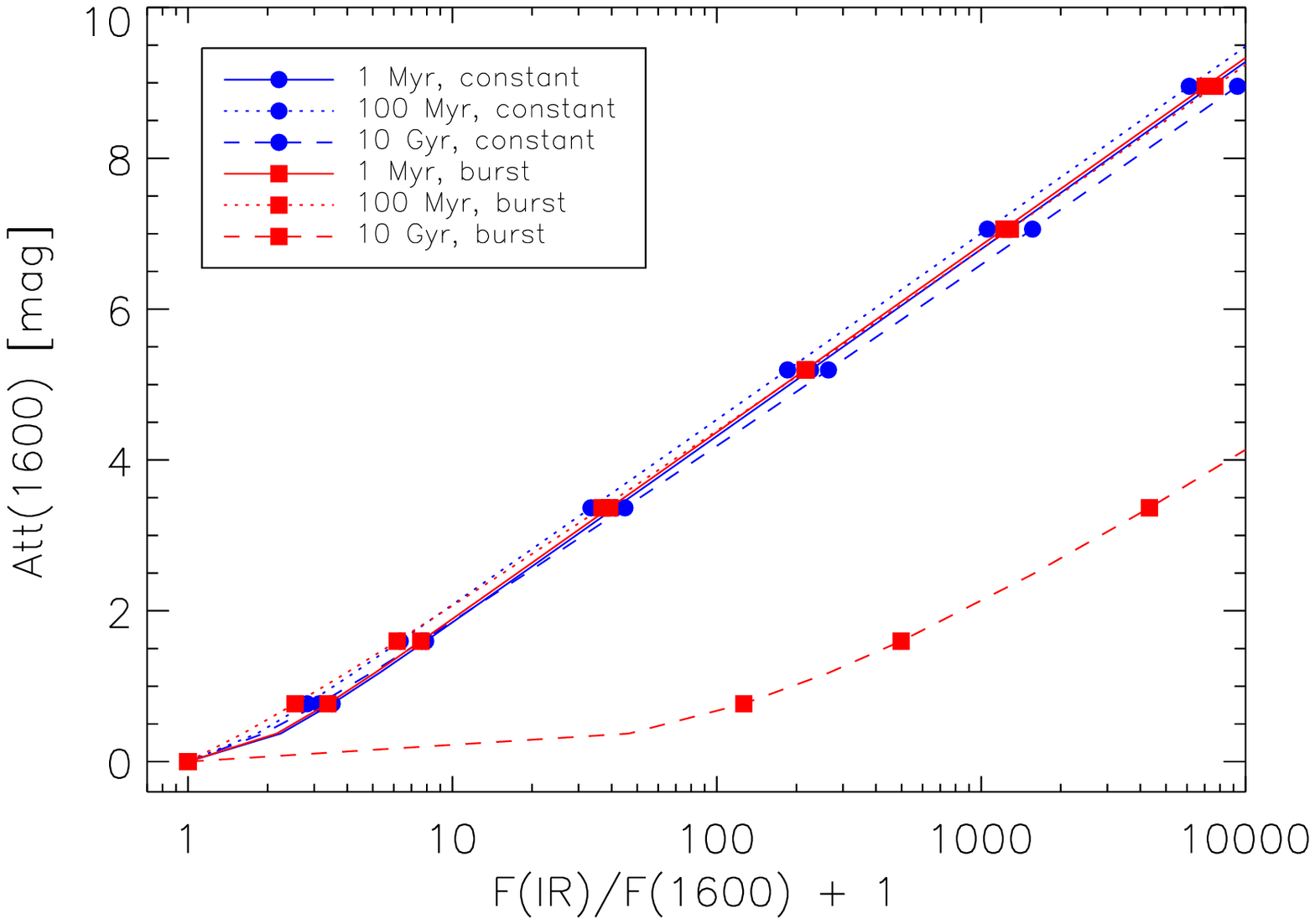}{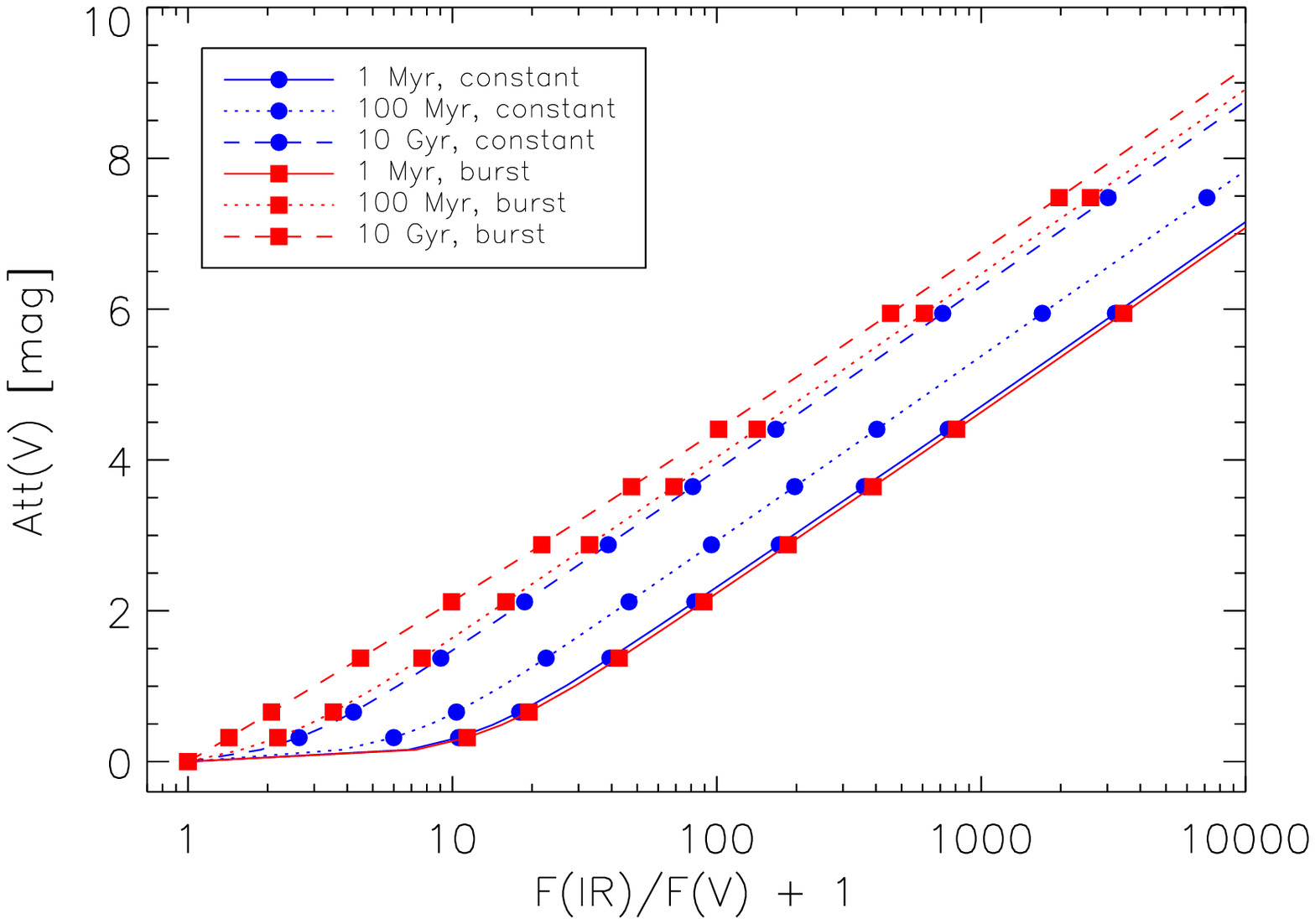} \\
\plottwo{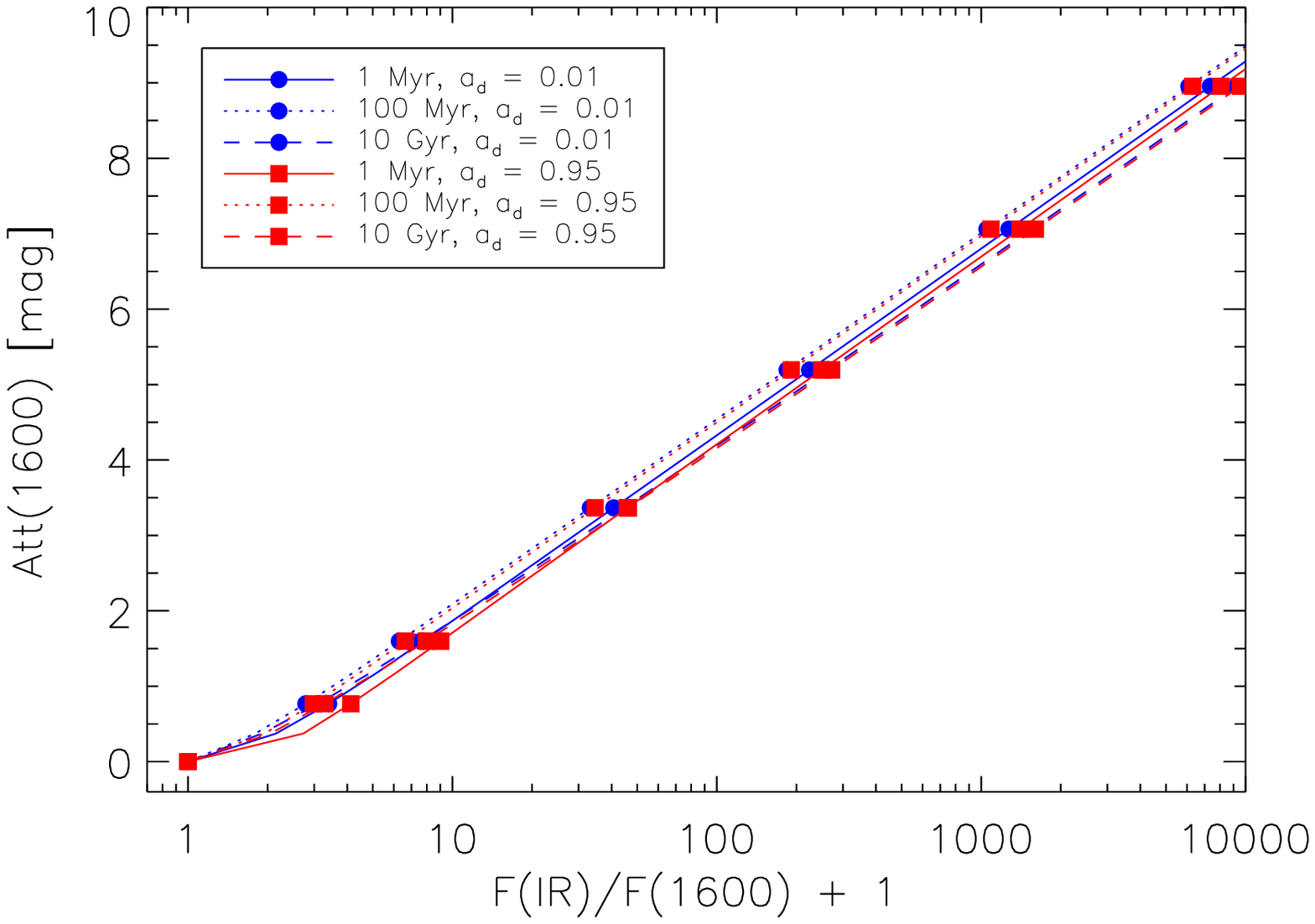}{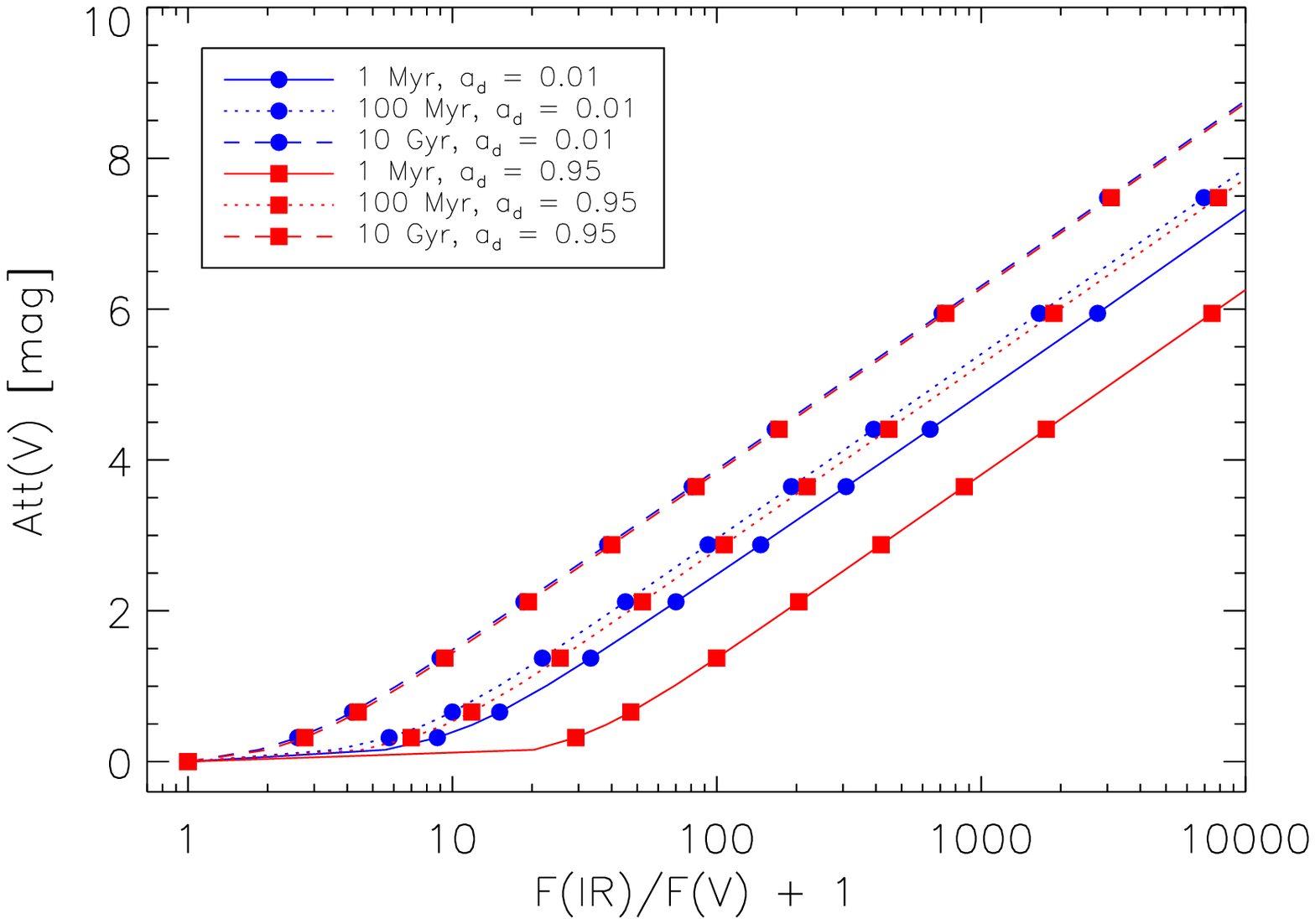}
\caption{The relationship between $F(IR)/F(\lambda)$ and
$Att(\lambda)$ is plotted for the 1600~\AA\ (left) and V (right)
bands.  The plots illustrate the dependence of the flux ratio
relationship on age and metallicity (a,b), constant versus burst star
formation (c,d), or $a_d$ value (e,f).  Only the DIRTY dust model run
with a homogeneous distribution, Milky Way-type dust, and a shell
geometry is shown for simplicity.  The other DIRTY model runs show
similar behaviors.  The stellar population parameters are nominally
solar metallicity (met = 0.0), constant star formation, and $a_d =
0.25$ except when one of these parameters was being varied.
In this figure, we have represent the stellar metallicity by $met =
\log(Z/Z_{\sun})$.
\label{fig_fr_all}} 
\end{figure}

\begin{figure}[tbp]
\plottwo{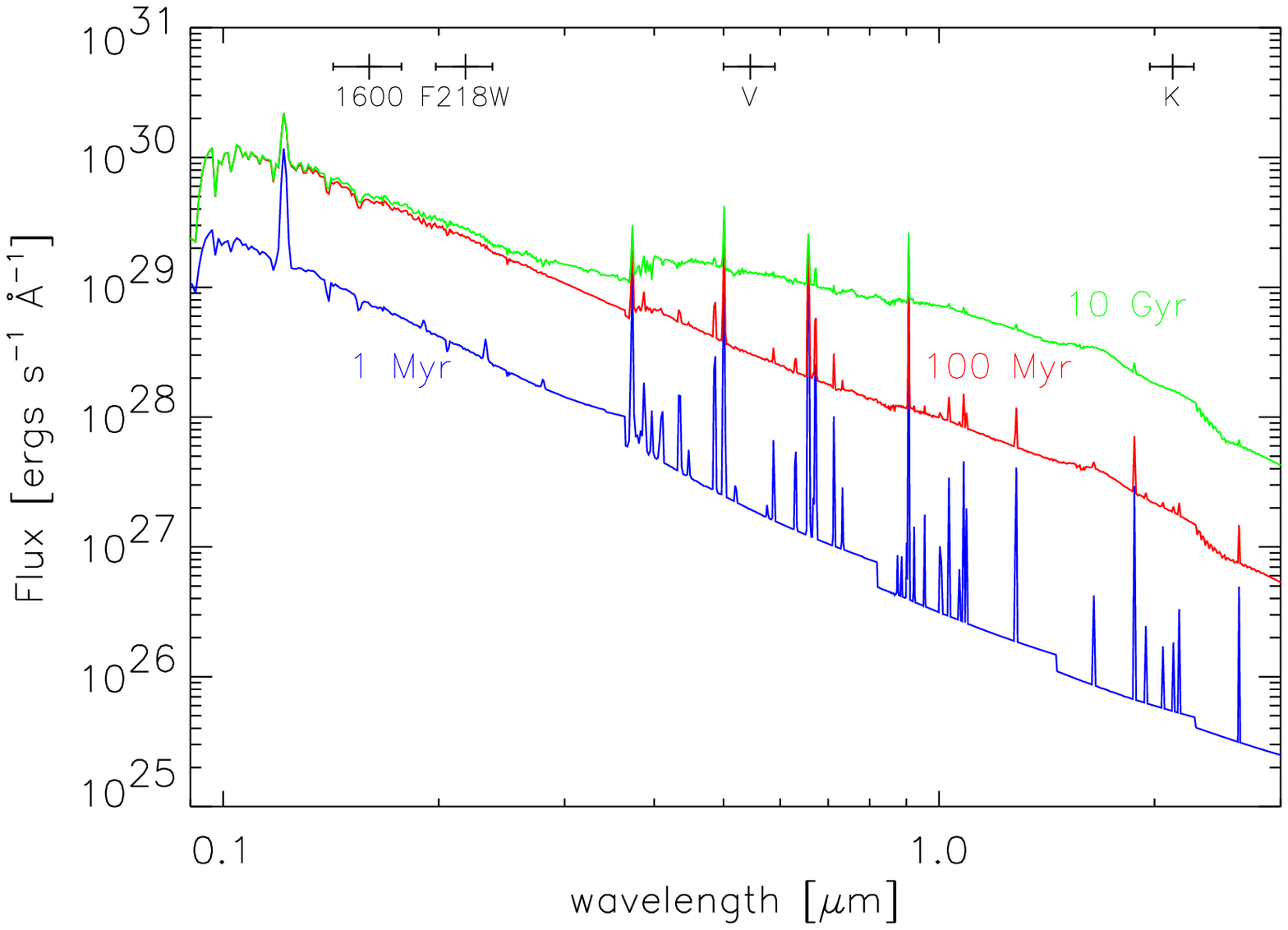}{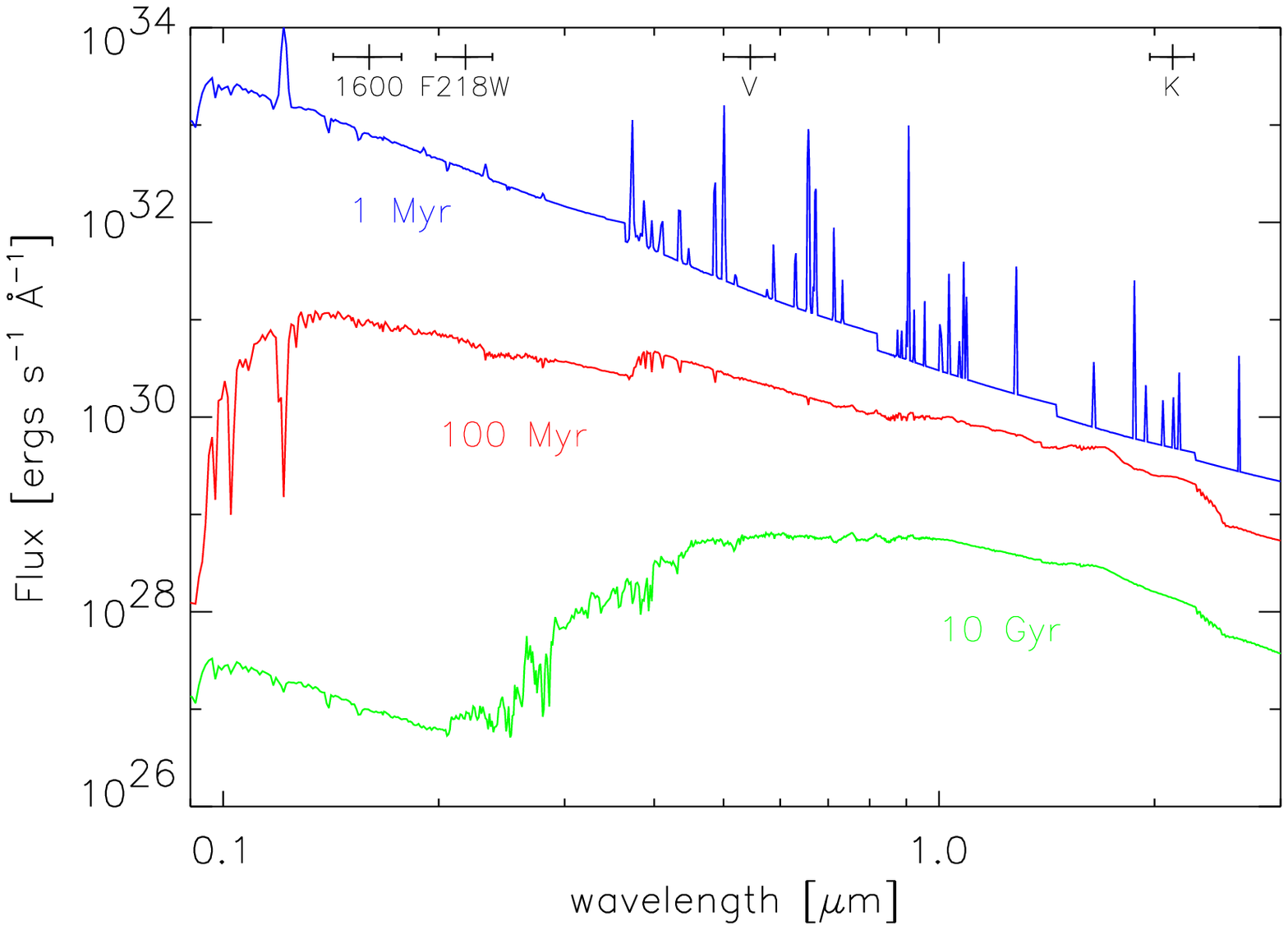} \\
\caption{Example PEGASE SEDs for constant (a) and burst (b) stellar
populations are plotted.  The constant star formation SEDs have a star
formation rate of 1 $M_{\sun}$ per year.  The burst star formation
SEDs have a total mass of $1 \times 10^{6}~M_{\sun}$.  The stellar
populations have solar metallicity (met = 0.0) and $a_d = 0.0$.  The
positions and widths of the four photometric shown in
Figs.~\ref{fig_fr_one} and \ref{fig_fr_all} are marked in these
plots. \label{fig_ses_mods}} 
\end{figure}

The dependence of the $F(IR)/F(\lambda)$ versus $Att(\lambda)$
relationship can be sensitive to the shape of the intrinsic SED.
Example SEDs for solar metallicity stellar populations are
given in Fig.~\ref{fig_ses_mods}.  The dependence of
$F(IR)/F(\lambda)$ on $Att(\lambda)$ is illustrated in
Figure~\ref{fig_fr_all} which shows the dependence of the flux ratio
relationship for the 1600 and V bands on age, metallicity, star
formation rate, and $a_d$ value.  In the 1600 band, the
relationship is quite similar for most choices of the above parameters
except for old burst stellar populations (Fig.~\ref{fig_fr_all}c).
Our calibration of $F(IR)/F(1600)$ versus $Att(1600)$ is
indistinguishable from that presented in \cite{meu99} after correcting
for $\sim$30\% difference between $F_{FIR}$ \citep{hel88} and
$F(IR)$ as $F_{FIR}$ does not include the hotter dust detected in the
mid-IR.  In the V band, the relationship is dependent, in decreasing
order of dependence, on age, burst versus constant star formation,
metallicity, and value of $a_d$.  The qualitative dependence of other
UV bands ($\lambda < 3000$~\AA) is similar to that seen for the
1600~\AA\ band.  The behavior of optical and near-infrared bands is
similar to that of the V band with the increasing dependence on the
above parameters as $\lambda$ increases.

\begin{figure}[tbp]
\plottwo{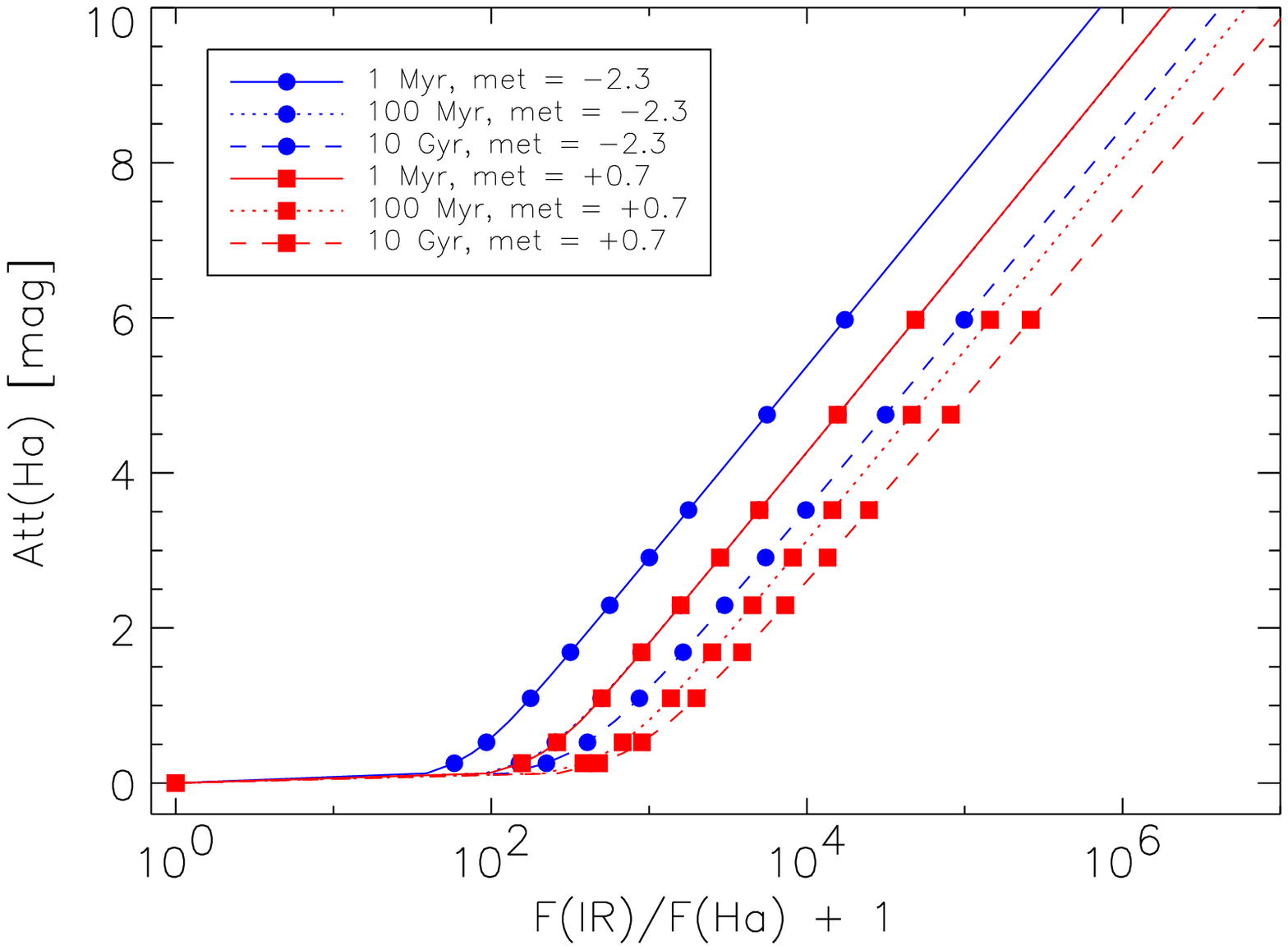}{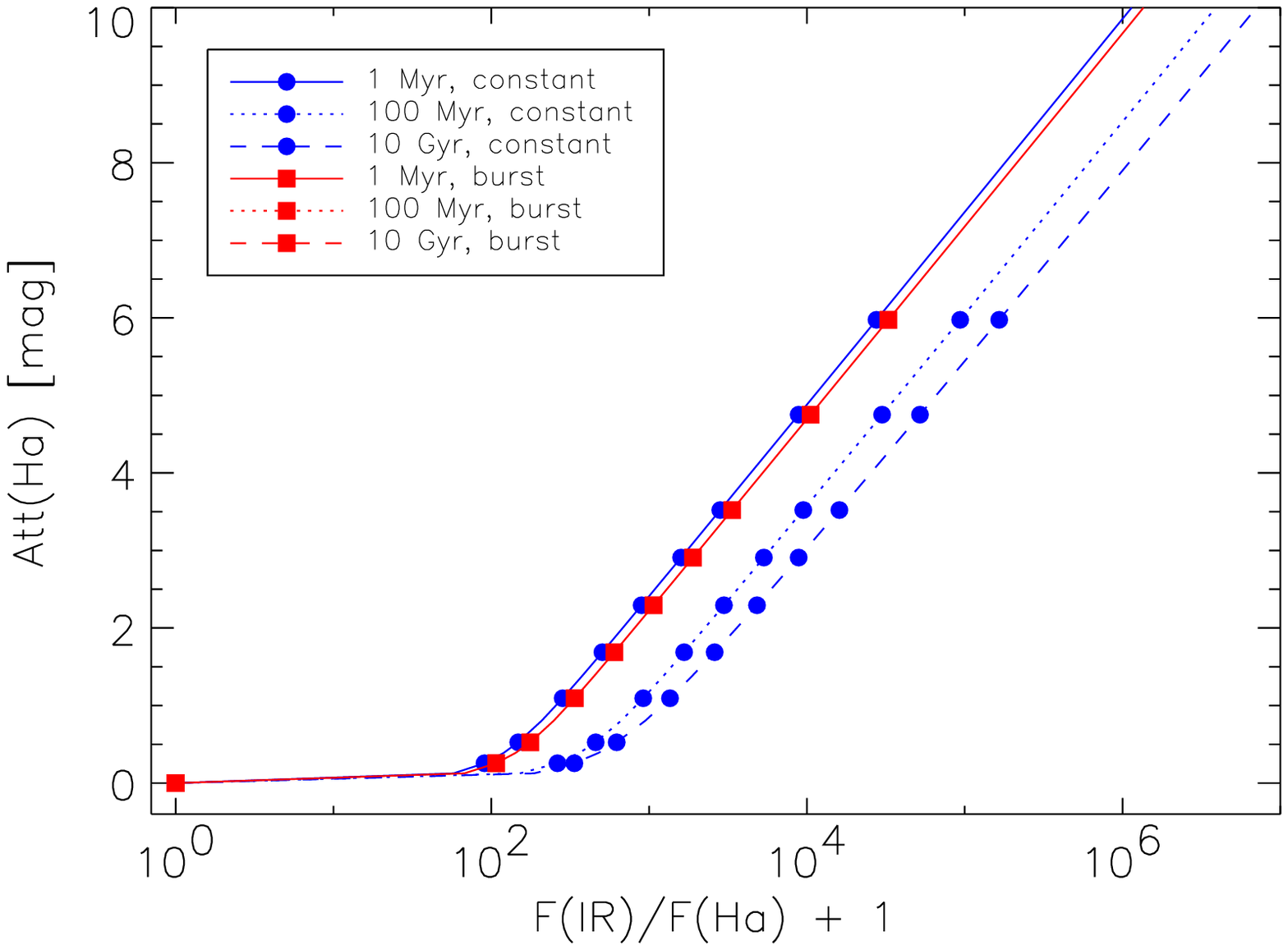} \\
\epsscale{0.5}
\plotone{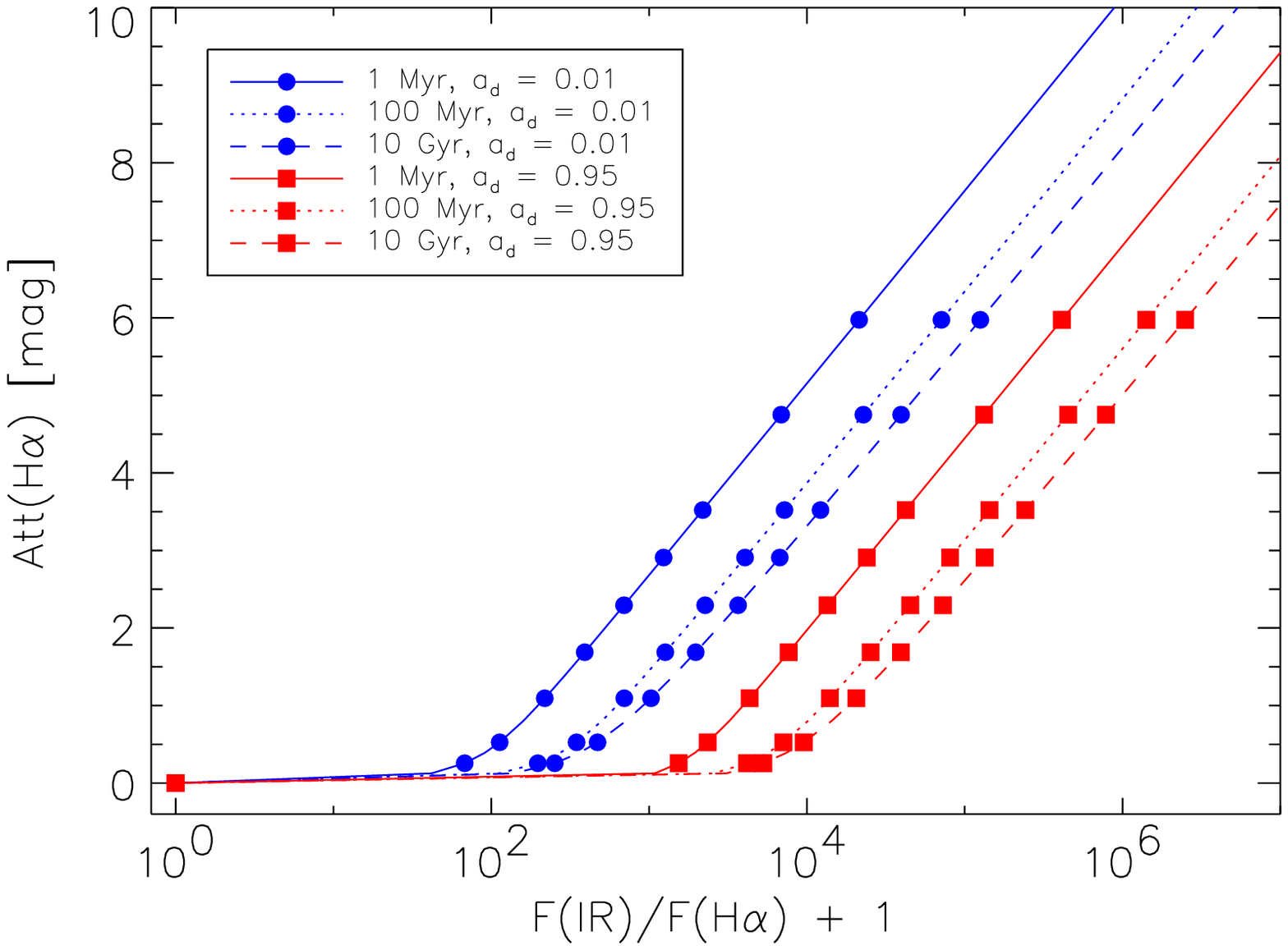}
\epsscale{1.0}
\caption{The relationship between $F(IR)/F(H\alpha)$ and
$Att(H\alpha)$ is plotted to show the dependence on age, metallicity
(a), burst versus constant star formation (b), and $a_d$ value (c).
Only the DIRTY dust model run with a homogeneous distribution, Milky
Way-type dust, and a shell geometry is shown for simplicity.  The
other DIRTY model runs show similar behaviors.  The lack of curves for
100 Myr and 10 Gyr burst stellar populations in plot (b) is due to the
lack of $H\alpha$ emission for these stellar populations
(Fig.~\ref{fig_ses_mods}).
\label{fig_fr_ha}}
\end{figure}

The behavior of emission lines is similar to that seen for the V band,
but has notable differences.  Figure~\ref{fig_fr_ha} gives the
relationship between $F(IR)/F(H\alpha)$ and $Att(H\alpha)$ for the
same parameters plotted in Fig.~\ref{fig_fr_all}.  One obvious
difference between the V band and H$\alpha$ emission line is that the
behavior with age is reversed.  In particular, the H$\alpha$ emission
line is {\em very} sensitive to the value of $a_d$ since the strength
of H$\alpha$ is directly proportional to $(1 - a_d)$.

The behavior of the flux ratio versus $Att(\lambda)$ relationship can
be qualitatively explained fairly easily.  The general shape of the
curves (see Fig.~\ref{fig_fr_one}) is seen to be non-linear
versus $F(IR)/F(\lambda) + 1$ below $Att(\lambda) \sim 1.5$ and nearly
linear versus $\log[F(IR)/F(\lambda) + 1]$ above $Att(\lambda)
\sim 1.5$. The non-linearity of the curve is due to changing
relationship between the effective wavelength of $F(IR)$ energy
absorption and that of $Att(\lambda)$.  The linear portion of the
curve is in the realm where $F(IR)$ is changing slowly (most of the
galaxy's luminosity is now being emitted in the IR), but $F(\lambda)$
continues to decrease due to the steady increase of $Att(\lambda)$.
Thus, above $Att(\lambda) \sim 1.5$ the curves for all wavelengths
have the same slope but different offsets reflecting the contributions
different wavelengths make to $F(IR)$.  Below $Att(\lambda) \sim 1.5$,
the effective wavelength of the 1600 and F218W bands is similar to
that of the $F(IR)$ energy absorption resulting in a nearly linear
relationship.  This is not the case for the V and K bands where their
effective wavelengths are much larger than that of the $F(IR)$ energy
absorption and, therfore, the V and K bands have non-linear
relationships below $Att(\lambda) \sim 1.5$.  The different
behaviors of the star/gas/dust geometry relationships (see
Fig.~\ref{fig_fr_one}) is due to the presence of stars outside the
dust in the cloudy geometry and the lack of external stars in the
shell and dusty geometries. 

The behavior of the relationship for different stellar populations
(Fig.~\ref{fig_fr_all} \& \ref{fig_fr_ha}) can be easily explained
using the same arguments used above.  The invariance of the
relationship for the 1600 band is a reflection of the dominance of the
UV in the $F(IR)$ energy absorption.  The only time when the 1600
relationship is not invariant is for old burst stellar populations
where the lack of significant UV flux means that the optical dominates
the $F(IR)$ energy absorption (Fig.~\ref{fig_ses_mods}b).  This is
confirmed by the linear behavior over the entire $Att(\lambda)$ range
of the V band curves for old stellar populations
(Fig.~\ref{fig_fr_all}d).  The separation of the curves in the V band
is the result of the different contributions the V band flux makes to
the $F(IR)$ absorbed energy for different stellar populations.  The
older the stellar population, the more the optical contributes to the
$F(IR)$ and, thus, the more linear the V band relationship is below
$Att(V) \sim 1.5$.

\subsection{Fits to the Relationships}

In order to use this method, we have fit the relationship between
$F(IR)/F(\lambda)$ and $Att(\lambda)$ for combinations of stellar age,
metallicity, burst or constant star formation, and values of $a_d$.
We chose to fit the combination of the dusty/shell geometry curves.
However, this does not limit the use of our fits in the UV since the
cloudy 
geometry curves follow the dusty/shell geometry curves.  This does
limit the use of our fits for wavelengths longer than $\sim$3500~\AA\
to cases where the dominant stellar sources are embedded in the dust
such as starburst galaxies.  The curvature of the relationship at
$Att(\lambda) \sim 1$ required us to use a combination of a 3rd order
polynomial for $Att(\lambda) < 1.75$ and a 2nd order polynomial for
$Att(\lambda) > 1$.  As a result the fit is:
\begin{equation}
\label{eq_fit}
Att(\lambda) = \left\{ 
   \begin{array}{ll} 
   A(x) & x < x_1 \\ 
   w(x)A(x) + (1 - w(x))B(x) & x_1 < x < x_2 \\ 
   B(x) & x > x_2 \\ 
   \end{array} \right.
\end{equation}
where
\begin{eqnarray*}
x & = & F(IR)/F(\lambda), \\ 
x_1 & = & x[Att(\lambda) = 1] \\ 
x_2 & = & x[Att(\lambda) = 1.75] \\ 
A(x) & = & a_1 + b_1x + c_1x^2 + d_1x^3, \\ 
B(x) & = & a_2 + b_2(\log x) + c_2(\log x)^2, \mbox{ and} \\ 
w(x) & = & (x_2 - x)/(x_2 - x_1).
\end{eqnarray*}
For each curve fit with equation~\ref{eq_fit}, 9 numbers result; 4
coefficients for $A(x)$, 3 coefficients for $B(x)$, the
$F(IR)/F(\lambda)$ values where $Att(\lambda) = 1$ and $1.75$ ($x_1$
and $x_2$).  Computing the
$Att(\lambda)$ value corresponding to a particular value of
$F(IR)/F(\lambda)$ then involves specifying the stellar age,
metallicity, star formation type, and value of $a_d$ which specify the
appropriate fit coefficients to use.  The parameters of these fits are
available from the lead author as well as an IDL function which
implements the calibration.

\section{Comparison with Radio Method \label{sec_compare}}

While the flux ratio method is relatively simple and makes sense
qualitatively, to be truly convincing, we need an independent method
for determining the attenuation for comparison.  Fortunately, radio
observations combined with measured hydrogen emission line fluxes
allows just such a test.  The radio method \citep{con92} is based on
the measurement of the free-free radio flux from \ion{H}{2} regions
and the assumption of Case B recombination \citep{ost89}.  From the
thermal flux, the number of Lyman continuum photons absorbed by the
gas can be calculated and, thus, the intrinsic fluxes of the hydrogen
emission lines.  Comparison of the intrinsic and observed line fluxes
gives the attenuation at the emission line wavelength.  The major
source of uncertainty in the radio method is that radio observations
contain both thermal (free-free) and nonthermal (synchrotron)
components.  For example, approximately a quarter of the flux measured
at 4.85 GHz has a thermal origin.  The decomposition of the measured
radio flux into thermal and nonthermal components imparts a factor of
two uncertainty in the resulting thermal flux \citep{con92}.

Unfortunately, determining attenuations using the flux ratio method is
the most uncertain for hydrogen emission line fluxes.  This is due to
the lack of knowledge of the value of $a_d$, the fraction of Lyman
continuum photons absorbed by dust (Fig.~\ref{fig_fr_ha}).  We can
take guidance from the work done by \citet{deg92} on six Large
Magellanic Cloud \ion{H}{2} regions.  She found that $a_d$ ranges from
0.21 -- 0.55 using the approximation of \citet{pet72}.  We will use
this range of $a_d$ values in the calculations below.

To do this comparison, we need galaxies which have hydrogen emission
line fluxes, infrared, and radio observations.  In the IUE sample of
starburst galaxies \citep{kin93}, there are 10 galaxies with Balmer
emission line \citep{sto94, mcq95}, IRAS \citep{cal95}, and 4.85 GHz
observations \citep{gre91,wri94,wri95,wri96}.  The 10 galaxies are NGC
1313, 1569, 1614, 3256, 4194, 5236, 5253, 6052, 7552, \& 7714.  The
emission lines were measured in a $10\arcsec \times 20\arcsec$
aperture which was usually large enough to include the entire
starburst region but not the entire galaxy.  While the IRAS and 4.85
GHz observations usually encompass the entire galaxy, the majority of
the IRAS and radio flux emerges from the starburst region which should
minimize the importance of the aperture mismatch \citep{cal95}.

\begin{figure}[tbp]
\begin{center}
\epsscale{0.5}
\plotone{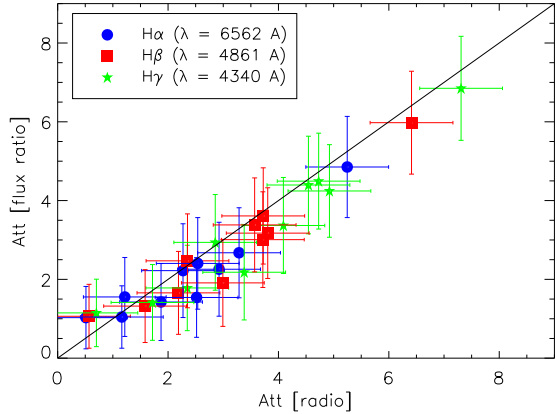}
\epsscale{1.0}
\end{center}
\caption{The attenuations of the H$\alpha$, H$\beta$, and H$\gamma$
emission lines calculated from the flux ratio method are plotted
versus those computed using the radio method.  The solid line
represents the case where $Att[\mbox{radio}] = Att[\mbox{flux
ratio}]$. \label{fig_compare}}
\end{figure}

Figure~\ref{fig_compare} shows the comparison between the attenuations
suffered by the H$\alpha$, H$\beta$, and H$\gamma$ emission lines in
the 10 galaxies as calculated from the flux ratio method and the radio
method.  While the measurements of each galaxy's three Balmer emission
lines are related (through Case B recombination theory), plotting all
three reduces the observational uncertainty due to the emission line
flux measurements and increases the range of attenuations tested.  For
the radio method, we calculated the intrinsic emission line strengths
using eqs.~3 \& 5 of \citet{con92} assuming a $T_e = 10^4~K$ and
Table~4.2 of \citet{ost89}.  The attenuations were then easily
calculated from the intrinsic and observed emission line fluxes.

For flux ratio method, the $F(IR)$ flux was computed by integrating
each galaxy's 8 to 1000 $\micron$ SED after extrapolating the IRAS
fluxes to longer wavelengths using a modified black body (dust
emissivity $\propto \lambda^{-1}$).  The temperature and flux level of
the modified black body were determined from the IRAS 60 and
100~$\micron$ fluxes.  ISO observations of starburst galaxies support
the use of a single temperature for the large dust grain emission
\citep{kru98}.  The 10 galaxies' $F(IR)$ fluxes were 1.6 to 2.5 times
larger than their FIR fluxes (as defined by \citet{hel88}) due to our
inclusion of the mid-infrared hot, small dust grains.  We assumed the
10 galaxies were undergoing 
constant star formation and used their measured metallicities
\citep{cal95} for the calculation of their attenuations from their
measured infrared to emission line flux ratios.  The error bars in
Fig.~\ref{fig_compare} for the flux ratio method reflect the range of
attenuations possible, assuming the galaxy age is between 1 Myr and 10
Gyr and $a_d$ values between 0.21 and 0.55.

The attenuations calculated for the two methods agree well within
their associated uncertainties.  This gives confidence that the flux
ratio method for calculating attenuations is valid.  Of course, this
conclusion would be strengthened with a larger sample of galaxies and
observations with similar apertures at optical, infrared, and radio
wavelengths.  Such infrared observations will become possible with the
launch of SIRTF.

\section{Application to Individual Galaxies \label{sec_app}}

The application of the flux ratio method to determining the UV
attenuations of individual galaxies is straightforward.  Due to the
insensitivity in the UV of this method to the star, gas, or dust
parameters (Figs.~\ref{fig_fr_one}a,b \& \ref{fig_fr_all}a,c,e), the
observed $F(IR)/F(UV)$ is directly related to $Att(UV)$.  This is not
the case for optical and near-IR wavelengths where this method is
sensitive to the intrinsic SED shape (Figs.~\ref{fig_fr_all}b,d,f)
and, to a lesser extent, the geometry of the star, gas and dust
(Fig.~\ref{fig_fr_one}c,d).

In order to construct the full UV through near-IR attenuation curve
for a galaxy, an iterative procedure must be followed.  The steps of
the iterative procedure are:
\begin{enumerate}
\item Assume an intrinsic SED shape (stellar age, metallicity, star
formation type, and $a_d$ value),
\item Construct a candidate attenuation curve using the observed
UV-NIR $F(IR)/F(\lambda)$ and our calibration of $Att(\lambda)$ versus
$F(IR)/F(\lambda)$,
\item Deredden the observed UV-NIR SED with the candidate attenuation
curve,
\item Compare the dereddened SED (step 3) with the assumed SED (step 1),
\item Repeat steps 1-4 to find the attenuation curve which produces
the best match between the dereddened SED and the assumed SED.
\end{enumerate}

We attempted to apply this iterative method to 10 starburst galaxies
listed in the previous section as they have UV, optical,
near-infrared, infrared, and radio observations.  We were unable to
find fits which would simultaneously fit the UV/optical/NIR continuum
and the H$\alpha$ emission attenuations derived from the ratio
observations.  To do the fitting we used the measured metallicities of
the galaxies and allowed the galaxy's age and type of star formation
as well as the the value of $a_d$ to vary.  The fact that we could not
find fits to any of the 10 galaxies is an indication that at least two
stellar populations are contributing to the observed SED.  But, the
correlation between the radio and flux ratio H$\alpha$ attenuations is
strong evidence that only one of these stellar populations is ionizing
the gas and is the main source for the dust heating (see
Fig.~\ref{fig_compare}).  This stellar population is likely the
starburst and the other stellar population is likely that of the
underlying galaxy.  The existence of two stellar populations, each
with different stellar parameters and attenuation curves, complicates
the fitting to the point where the number of free parameters can
exceed the number of observed data points.  This illustrates one of
the main difficulties of applying the flux ratio method.  The
calibration of the flux ratio method is based on the assumption that
there is a single stellar population responsible for the UV-NIR
continuum and IR dust emission.  When a second stellar population
contributes to the continuum or IR dust emission, applying the flux
ratio method will become more difficult.

\begin{figure}[tbp]
\begin{center}
\epsscale{0.5}
\plotone{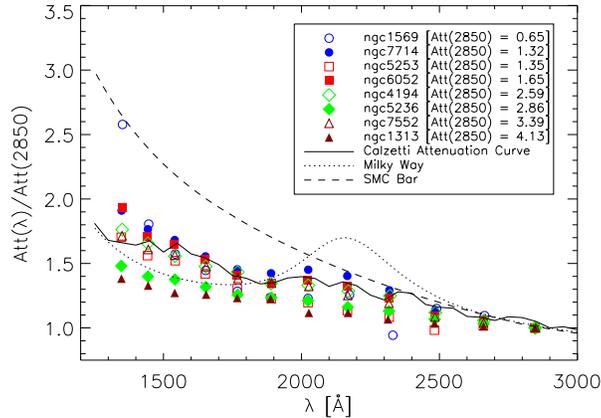}
\epsscale{1.0}
\end{center}
\caption{The UV attenuation curves normalized to $Att(2850)$ for 8 of
the 10 starburst galaxies using in the previous section are plotted.
Two galaxies were excluded as they did not have near-UV data.  The
Calzetti attenuation curve for starburst galaxies \citep{cal97}, the
Milky Way $R_V = 3.1$ extinction curve \citep{car89}, and SMC Bar
extinction curve
\citep{gor98} are plotted for comparison.  \label{fig_uv_curves}}
\end{figure}

While we cannot determine the UV-NIR attenuation curves for the 10
starburst galaxies, we can determine their UV attenuation curves for
the following reason.  The identification of the stellar population
heating the dust as the same population ionizing the gas leads us to
conclude that the same population also emits the majority of the
galaxies' UV continua since UV photons are the main source of dust
heating for starburst galaxies (see \S\ref{sec_compare}).
Figure~\ref{fig_uv_curves} gives the attenuation curves for 8 of the
10 starburst galaxies used in the previous section.  The other 2
galaxies were excluded as they did not have near UV data.  All of the
curves lack a substantial 2175~\AA\ bump in agreement with previous
work \citep{cal94,gor97}.  There is also a trend towards steeper
attenuation curves as $Att(2850)$ decreases which is the behavior
predicted by \citet{wit99}.  The curves are similar to the ``Calzetti
attenuation curve'' \citep{cal94,cal97} derived for the IUE sample of
galaxies.  As the 
Calzetti attenuation curve is an average, the scatter of our
individual curves is likely to be real.  While the 8 galaxies in
Fig.~\ref{fig_uv_curves} were included in the \citet{cal97} work, the
method used to derive the Calzetti 
attenuation curve was quite different from the flux ratio method.
This is further evidence that the flux ratio method can determine the
attenuation curves of starburst galaxies.

\section{Discussion \label{sec_discuss}}

We have presented a method which uses the $F(IR)/F(\lambda)$ flux
ratio to determine $Att(\lambda)$ for individual starburst galaxies.
The major strengths of this method is that it is almost completely
independent of the type of dust (MW/SMC) or the local distribution of
dust (homogeneous/clumpy), and is only weakly dependent on the global
distribution of stars and dust (presence/lack of stars outside dust).
In the ultraviolet, this method is independent of the intrinsic
stellar SED except for the case of very old burst populations.  In the
optical/near-IR, this method is dependent on the intrinsic stellar SED
shape.  The flux ratio method is not based on the properties of the
nebular emission (as is the radio method), but on the properties of
the stellar continuum and IR dust emission.  As a result, it is
applicable to any wavelength from the UV to near-IR and not just
wavelengths with hydrogen emission lines.

A major limitation of the flux ratio method is that the majority of
the observed UV through far-infrared flux must originate from a single
stellar population (either burst or constant star formation).  An
example of a case where the flux ratio method would not be applicable
would be a heavily embedded starburst in a galaxy with a second older,
less embedded stellar population.  At UV and IR wavelengths the
starburst would dominate, but at optical and near-IR wavelengths the
older population would dominate.  Another possible limitation is that
the measured infrared flux is assumed to be a direct measure of the
flux absorbed by the dust.  If the infrared radiation in not emitted
symmetrically (e.g., for non-symmetrically distributed dust which is
optical thick in the infrared), then the measured infrared flux will
not be a direct measure of the flux absorbed by the dust.  The
assumption that the infrared flux is a direct measure of the flux
absorbed by the dust is crucial to the accuracy flux ratio method.  It
is possible to account for these weaknesses by increasing the
complexity of the modeling by adding additional stellar populations
and/or complex dust geometries.  Such increases in the complexity of
the modeling will necessarily require more detailed spectral and
spatial observations as the number of model parameters increases.

For any starburst galaxy with UV and IR observations, the UV
attenuation curve can be calculated using the flux ratio method.
Starburst galaxies are likely to be the best case for applying the
flux ratio method as the intensity of the starburst greatly increases
the probability that the UV and IR flux originate from only the
starburst population.  If the parameters (age, metallicity, etc) of
the intrinsic SED shape can be determined and the contamination from
the underlying stellar population removed, then the attenuation of the
starburst galaxy can be determined not only for the UV, but also for
the optical and near-IR.

Thus, the flux ratio method seems very promising for determining
the dust attenuations of individual galaxies.  The easiest way to
ensure the basic assumptions of our calibration of the flux ratio
method are met is to take high spatial resolution observations of
starburst regions in nearby galaxies or integrated galaxy observations
of intense starburst galaxies at any distance.  This would ensure that
the UV through far-infrared flux originates from the starburst and not
the host galaxy.  Examples of these observations would be super star
clusters in nearby galaxies
\citep{cal97b} and observations of high-z starburst galaxies which have
been shown to be similar to local starbursts except more intense
\citep{hec98}.  Currently, both types of UV, optical, and near-IR
observations can and have been done, but the far-infrared observations
needed await SIRTF.  SIRTF will have the spatial resolution and
sensitivity to do both types of observations.

The ability to determine the UV dust attenuation curve for individual
starburst galaxies will facilitate the study of dust in different star
formation environments.  The traditional explanation for the
differences seen in the dust extinction between the Milky Way, LMC,
and SMC has been that the different metallicities of the three
galaxies lead to different dust grains.  Work on starburst galaxies
with metallicities between 0.1 and 2 times solar which found most of
these galaxies possess dust which lacks a 2175~\AA\ bump \citep{cal94,
gor97} seriously called this explanation into question.  Subsequent
work on the extinction curves in both the SMC \citep{gor98} and LMC
\citep{mis99} found that the extinction curves toward star forming
regions in both galaxies were systematically different than those
toward more quiescent regions.  These results imply that dust near
sites of active star formation is different due to processing
\citep{gor97} of existing dust or formation of new dust
\citep{dwe98}.  The processing interpretation is 
supported by recent work in the Milky Way along low density sightlines
toward the Galactic Center \citep{cla99}.  This work found that
sightlines which show evidence of processing (probed by N(Ca II)/N(Na
I)) have weaker 2175~\AA\ bumps and stronger far-UV extinctions than
most other Milky Way sightlines \citep{car89}.  The actual processing
mechanism is not simple as the dust towards the most intense star
formation in the LMC (30 Dor) has a weak 2175~\AA\ bump, but the dust
towards the most intense star formation in the SMC, which has only
10\% the strength of 30 Dor, has no 2175~\AA\ bump.  In order to
completely characterize the dust near starbursts, attenuation curves
for a large sample of starbursts galaxies with a range of metallicity,
dust content, and starburst strength are needed.

In conjunction with investigating the impact environment has on dust
properties, the ability to determine individual starburst galaxy
attenuation curves will simplify the study of the starburst
phenomenon.  By being able to remove the effects of dust accurately,
the age and strength of starburst galaxies and regions in galaxies can
be determined with confidence.  In the realm of high redshift
starburst galaxies ($z > 2.5$), the ability to determine the dust
attenuation of individual galaxies will arrive with the advent of deep
SIRTF/MIPS imaging of fields with existing rest-frame UV imaging (eg.,
Hubble Deep Fields).  The currently large uncertainty on the global
star formation history of the universe due to the effects of dust on
starburst galaxies will be greatly reduced \citep{mad98,pet98,ste99}.

\acknowledgements

This work benefited from discussions with Daniela Calzetti and
Gerhardt Meurer.  Support for this work was provided by NASA through
LTSAP grant NAG5-7933 and archival grant AR-08002.01-96A from the
Space Telescope Science Institute, which is operated by AURA, Inc.,
under NASA contract NAS5-26555.

\end{document}